\documentclass[11pt,a4paper]{article}
\usepackage{jheppub}
\pdfoutput=1
\usepackage[utf8]{inputenc}
\usepackage{amsfonts,amsmath,amssymb}
\usepackage{graphicx}
\usepackage{multirow}
\usepackage{verbatim}
\usepackage{bm}
\usepackage[textsize=scriptsize,backgroundcolor=red!70,linecolor=red]{todonotes}
\usepackage{paralist}
\usepackage{subfig}
\usepackage[normalem]{ulem}
\usepackage{csquotes}
\usepackage{adjustbox}
\usepackage{orcidlink}

\newcommand{\ii}{\mathrm{i}}
\newcommand{\ee}{\mathrm{e}}

\newcommand{\abs}[1]{\big\lvert #1 \big\rvert}

\newcommand{\ud}{\ensuremath{\mathrm{d}}}

\newcommand{\chisq}{\ensuremath{\chi^{2}}}

\newcommand{\porpb}{\ensuremath{\overset{\scriptscriptstyle \left(-\right)}{p}}}
\newcommand{\fform}{\ensuremath{\mathcal{F}}}
\newcommand{\bvec}[1]{\ensuremath{\bm #1}}
\newcommand{\dof}{\ensuremath{\text{d.o.f}}}
\newcommand{\apr}{\ensuremath{\alpha^\prime_P}}
\renewcommand{\Im}{\ensuremath{\mathfrak{Im}}}

\title{A multi-channel $U$-Matrix model of hadron interaction at high energy}
 
\author{Rami Oueslati \orcidlink{0000-0003-2569-4056}}
\emailAdd{rami.oueslati@uliege.be}

\affiliation{Space sciences, Technologies and Astrophysics Research (STAR) Institute, Université de Liège, Bât.~B5a, 4000 Liège,  Belgium}
\date{\today}

\abstract{The present phenomenological study investigates  a multi-channel model of 
high-energy hadron interactions by considering a full parton configurations space and the $U$-matrix unitarisation scheme of the elastic amplitude, comparing it to the two-channel model, and  examining the consequences of up-to-date high-energy collider data on the best fits to various hadronic observables  in $pp$ and $p \bar{p}$ collisions. The findings of this study reveal that the data are well-fitted with the multi-channel model and that the difference compared to the two-channel one is negligible. Of particular significance is the observation that the $U$-matrix unitarisation is likely incompatible with uncorrelated pomeron exchange, as suggested by the equivalence between the $U$-matrix multi-channel and eikonal two-channel descriptions. Based on our best fit, predictions for the $\rho$ parameter, the double diffractive cross-section, and the elastic differential cross-section are provided. We shed light on the effect of taking into account a multi-channel model on present and future cosmic ray data.}


\begin{document}

\maketitle
\section{\label{sec:intro}Introduction}

When ultra-high energy cosmic ray particles first hit the Earth's atmosphere, several additional interactions take place. These interactions lead to particle multiplication and decay processes, which collectively result in a cascade of secondary particles known as an extensive air shower (EAS). In fact, observing these air showers is the only means of detecting high-energy cosmic ray particles. The development of air showers is dependent on hadronic cross sections and particle production characteristics in hadronic interactions.

However, it is worth noting that there is still much to be discovered about the evolution of the total and elastic cross sections in hadron-hadron collisions as functions of the center of mass energy as well as the characteristics of multiparticle production in these interactions. It is an interesting line of
research given its phenomenological implications. 

Indeed, estimating the features of hadronic interactions at LHC energies is crucial not only for modelling the background while looking for potential manifestations of new physics but also for the interpretation of the existing (and future) cosmic ray data, which relies on theoretical assumptions that describe these interactions.

As a matter of fact, despite being a well-known and experimentally supported theory of strong interactions, Quantum chromodynamics (QCD) can only currently predict processes involving large momentum transfer. Furthermore, the bulk properties of multiparticle production, which are required for air shower simulation, are still not calculable. Therefore, in order to create models for hadronic interactions that describe various particle generation processes, it is necessary to make further simplifying hypotheses in conjunction with phenomenological models that essentially consist of perturbative QCD (pQCD) predictions and phenomenological fits to experimental hadron spectra, which in turn are based on  fundamental principles of quantum field theory – such as unitarity, analyticity and crossing, along with empirical parametrizations 
\cite{DENTERRIA201198}. Certainly, it is crucial to validate these assumptions, constrain the parametrizations, and fine-tune the parameters using accelerator data comparisons.

For instance, in a prior study \cite{Vanthieghem_2021}, the hypothesis of  using two different unitarization schemes; the commonly employed eikonal as well as the U-Matrix, as unitarity constraint of the elastic amplitude was examined by looking into the effect of including recent collider data for total, elastic, inelastic and single diffractive cross sections in the framework of the two-channel model. The results showed nearly identical cross-sections, regardless of the unitarisation scheme adopted. Most importantly, it has been found that the single diffractive data are slightly better described with the U-matrix than with the eikonal one, in spite of the data used. Another hypothesis with regard to considering an infinite parton configurations space has been examined using the eikonal scheme \cite{Broilo2020}. We intend to investigate this hypothesis, but rather utilizing the U-matrix scheme.

It should be noted that the U-matrix scheme is not used as an alternative to phenomenologically studying hadronic interactions at high energy, but rather for physical reasons. To start with, the choice of the U-matrix scheme is motivated by the aforementioned result \cite{Vanthieghem_2021}.  Secondly, owing to the fact that correlations may emerge from the fluctuations of the hadrons in various configurations, which is a phenomenon closely connected to hadron diffraction \cite{Treleani_2008}, we then may infer that these hadron fluctuations might be increased through implementing a multi-channel model of high energy hadronic interactions using the U matrix scheme. We expect that it will produce a better description of the hadronic observables, compared with the eikonal scheme, within the multi-channel model. We also anticipate that it will provide better results within the multi-channel model than within the two-channel one.

This study has the following objectives. First of all, it will focus on testing the hypothesis of considering  an infinite parton configuration space and compare it to the two-channel counterpart. Based on our model, it also seeks to predict the double diffractive cross-section, the ratio of the real part to the imaginary part of the elastic amplitude, i.e., the $\rho$ parameter, and the elastic differential cross-section.
Finally, the impact of considering a multi-channel model on present and future cosmic ray data will be discussed.

The present paper is organised as follows. In section \ref{sec:MC GW} we will focus on  the theoretical framework of the diffractive excitation in the context of the multi-channel Good-Walker approach. In section \ref{sec:Exp mod}, an explicit model for the description of the elastic scattering amplitude as well as the treatment of the average number of interactions will be proposed. Moreover, the principal parameters of the model and data used will be highlighted. In section \ref{sec:result}, the results of the study will be presented and discussed.
In section \ref{sec:conclusions}, the conclusions will be given.

\section{\label{sec:MC GW} Diffraction and multi-channel Good-Walker approach}

\subsection{ \label{sec:Teo fram} Theoretical framework}

Hadrons are composite particles comprised of quarks and gluons which interact in a variety of ways during  hadron collisions. It is possible to relate these interactions to the total and elastic cross sections using a suitable theoretical framework. But the specific way to achieve it is still an open question. In fact, "Mini-jet" models \cite{Durand_1988} are thought to be a viable option, with total and elastic cross sections calculated using an eikonal formalism in terms of the quantity  $\langle n(b,s)\rangle$, representing the average number of elementary interactions at impact parameter b and c.m. energy $\sqrt{s}$. 

It should be emphasized that predictions made using the simple eikonal scheme in these Mini-jet models are insufficient. The fact that this kind of elastic amplitude unitarization scheme is inappropriate for a collision of composite objects like hadrons is already supported by the findings of a number of studies \cite{Martynov_2020, Troshin:2003wu, Troshin_2020} as well as certain (indirect) evidence. In fact, the eikonal unitarization scheme is a well-known technique for calculating the amplitude $X$, which meets some minimal s-channel unitarity constraints from the \enquote{non-unitary} amplitude $\chi$, as 

\begin{equation}
    X =  i (1 -\exp{ (i \chi) } )
    \label{eq:eiko}
\end{equation} It is based on the assumption that the impact parameter (the perpendicular distance between the trajectories of colliding particles) is much larger than the characteristic size of the interacting particles. Regarding the statistical nature of this scheme, in collisions at a fixed impact parameter and c.m. energy, the fluctuations in the number of interactions are just Poissonian in nature \cite{Boreskov:2005ee}. The statistically independent and identically distributed  interactions is equivalent that each exchange process is statistically equivalent and contributes equally to the overall scattering amplitude. This is equivalent to a sum of contributions derived from the multiple exchanges, emerging with even weights, which is described by the primary amplitude $\chi$. While the assumption of equal weights is a useful simplification, it may not always accurately reflect the underlying physics. In reality, the individual exchange processes may have different strengths or probabilities, which could affect the overall scattering amplitude. Accounting for such differences would require a more detailed and sophisticated treatment beyond the eikonal approximation. Mathematically, the eikonal approximation allows us to factorize the overall scattering $S$-matrix associated with the interaction into a product of individual scattering matrices. This approach is sometimes connected with the image of a rapid particle travelling virtually straight ahead in target media, \cite{Glauber}.

Furthermore, the eikonal approximation treats the hadrons as classical objects with fixed parton distributions. It assumes that during the interaction, the parton configurations remain frozen or unchanged. This approximation is valid when the timescale for the parton dynamics, such as radiation and absorption, is much longer than the timescale of the interaction itself \cite{Lipari2009}. The freezing of parton configurations in the eikonal approximation simplifies the calculations by considering the partons as fixed distributions and focusing on the overall scattering process rather than the detailed internal dynamics. However, it is important to note that the freezing of parton configurations is an approximation and may not capture all aspects of the parton dynamics accurately. In reality, partons can undergo radiation and absorption processes, leading to changes in their energy and momentum distributions. Due to these limitations in the eikonal approximation, all such multiple exchanges may not occur simultaneously and may be dependent on each other. This challenges the assumption of equal weights and Poissonian behavior in the summation of exchanges. Furthermore, the need for multiple exchanges arises to account for phenomena such as screening effects and additional inelastic processes. The prevalence of the eikonal scheme in Monte Carlo event generators, such as SIBYLL \cite{SIBYLL_2020} and QGSJET \cite{QGSJET}, prompts a reevaluation of its suitability for unitarizing the elastic amplitude in hadronic collisions. Continual assessment and refinement of theoretical frameworks and models are necessary to better capture the complexities of high-energy interactions.

The fluctuating structure of hadrons, which are composite particles made up of quarks and gluons bound together by the strong force, is thought to contribute to the process of diffractive excitation. The internal structure of hadrons is highly complex and dynamic, with quarks and gluons constantly interacting and creating temporary resonances within the hadron. During a high-energy collision between two hadrons, these resonances can be excited by the exchange of a pomeron, leading to diffractive excitation. The exact mechanism of this process is still an area of active research in particle physics. According to Good and Walker (GW) \cite{GW_1960}, inelastic diffraction occurs because an interacting hadron can be perceived as a superposition of several states that experience uneven absorptions. GW further depicted the diffractive excitation as the eigenstates of the scattering operator, which are utilized to describe the physical states.

In the same vein, Miettinen and Pumplin \cite{MP_1978} postulated that these \enquote{transmission eigenstates} can be recognized as distinct \enquote{configurations} of the parton elements contained within a hadron. It is necessary to have a general grasp of the entirety of these parton configurations in order to estimate inelastic diffraction within this theoretical framework, which seems to be a challenging task. One possible method of doing so consists in lessening the space of parton configurations to a finite dimensional space and explicitly creating a matrix transition operator. As an illustration of this approach \cite{Vanthieghem_2021}, we have taken into account the minimal scheme initially proposed by Gotsman, Levin, and Maor  (GLM) \cite{GLM_1999}  and combined the proton with one diffractive state. This is equivalent to a two-channel unitarisation scheme. Another illustration can be found in \cite{GLM_1999}, where GLM examined the case $N = 3$ in the eikonal scheme but discovered no appreciable improvement.

Therefore, in order to highlight the difference in the description of the hadronic observables between the models, the entirety of parton configurations as well as the scheme adopted should be taken into account. Practically speaking, an  $N$ channel scheme could be considered, but this would increase the number of parameters, which will affect the attainment of a reliable and realistic model in comparison with the physics that we aspire to describe.

An alternative approach assumed here is to map the space of the parton configurations into the real positive numbers. Various research papers in the field have already explored this approach \cite{Avsar_2007, Flensburg_2010, Flensburg:2008ag, Flensburg_2011, Flensburg_2012, Gustafson_2015, Bierlich:2016smv, Broilo2020} but all of them  with the eikonal scheme. However, since no single published study, to our knowledge, has estimated the inelastic diffraction within the GW approach by considering the entirety of parton configurations together with the U-matrix scheme, this study attempts to fill this gap, at least partially. The total cross-section and its various constituents (elastic, absorption, and diffraction) can be calculated as will be shown in the following section.

\subsection{ \label{sec:form} Formalism}
We adopt the multichannel formalism presented in \cite{Broilo2020, Gustafson_2015}, with a small modification to account for a full complex scattering amplitude. The starting point is the impact parameter space representation, where the hadronic observables, the total, elastic, single, and double diffractive scattering cross sections may be readily expressed  as :

\begin{subequations}
\begin{align}
  \sigma_{tot}(s)  &= 2 \int\! \ud^2 b\ \Im\left\{ X_{el}(s, b) \right\}\,;
  &\sigma_{el}(s)  &= \int\! \ud^2 b\ \abs{X_{el}(s, b) }^2\,; \\
  \sigma_{sd}(s)   &= 2\int\! \ud^2 b\ \abs{X_{sd}(s, b) }^2 \,;
  &\sigma_{dd}(s)  &= \int\! \ud^2 b\ \abs{X_{dd}(s, b) }^2
\end{align}
\label{eq:sigmas}
\end{subequations}

When a projectile $P$ collides with a target $T$, represented by the physical states  $|P\rangle$ and $|T\rangle$ respectively, we assume that both states can be diffracted onto various particle states $\{|A\rangle\}$ and $\{|B\rangle\}$ due to their substructure. The GW approach states that the initial state can be expressed as a sum over the eigenstates $\{|\Psi_i\rangle\}$ of the scattering operator $\hat{T}$, forming a complete set of normalized states. This gives us the initial state $|I\rangle$ as:
\begin{equation}
|I\rangle = |P,T\rangle = \sum_{ij}C_i^PC_j^T|\psi_i\psi_j\rangle
 \end{equation}

where $\hat{T}|\psi_i\psi_j\rangle = t_{ij}|\psi_i\psi_j\rangle$, with the eigenvalues  $t_{ij} = t_{ij}(b, s)$ depending implicitly on the projectile and target's specific configurations. 
The final state system can be described by
\begin{equation}
|F\rangle = \hat{T}|I\rangle = \sum_{i,j}C_i^PC_j^Tt_{ij}|\psi_i\psi_j\rangle    
\end{equation} leading to
\begin{equation}
\langle F|F\rangle = \sum_{i,j}|C_i^P|^2|C_j^T|^2|t_{ij}|^2    
= \sum_{i,j}P_i^PP_j^T|t_{ij}|^2 = \langle|t|^2\rangle_{P,T}
\end{equation} where we have identified $P_i^P = |C_i^P|^2$ and $P_j^T = |C_j^T|^2$ as configuration's  probability distributions for projectile and target respectively, and $\langle...\rangle_{P,T}$ refers to the mean value calculated across the different configurations present in both the projectile and the target.

The final state system can be expressed as a sum over the possible final states $\{|A,B\rangle\}$, which form a complete set of eigenstates, as:

\begin{equation}
|F\rangle = \sum_{A,B}|A,B\rangle = |P,T\rangle + \sum_{A\neq P}|A,T\rangle + \sum_{B\neq T}|P,B\rangle + \sum_{A\neq P,B\neq T}|A,B\rangle    
\end{equation} As a result, we can deduce that:
\begin{eqnarray}
\langle F|F\rangle & = & \sum_{A,B} \langle F|A,B \rangle \langle A,B | F \rangle \nonumber  \\
& = & |\langle P, T | F \rangle |^2 + \sum_{A \neq P} |\langle A, T | F \rangle |^2 + \sum_{B \neq T} |\langle P, B | F \rangle |^2 + \sum_{A \neq P; B \neq T} |\langle A, B | F \rangle|^2 \,\,.
\end{eqnarray} Furthermore, by using the fact that

\begin{eqnarray}
X_{el}(s, b) \equiv \langle P, T | F \rangle & = & \sum_{i,j} |C_i^P|^2 |C_j^T|^2 t_{ij} \equiv \langle t \rangle_{P,T} \,\,,\\
X_{sd}^P(s, b) \equiv \langle A, T | F \rangle|_{A \neq P} & = & \sum_{i,j} C_i^{*,A} C_i^P |C_j^T|^2 t_{ij} \,\,, \\
X_{sd}^T(s, b) \equiv \langle P, B | F \rangle|_{B \neq T} & = & \sum_{i,j} |C_i^P|^2 C_j^{*,B} C_j^T  t_{ij}\,\,,\\
X_{dd}(s, b) \equiv \langle A, B | F \rangle|_{A \neq P; B \neq T} & = & \sum_{i,j} C_i^{*,A} C_i^P C_j^{*,B} C_j^T  t_{ij}^*  t_{ij} \,\,,
\end{eqnarray}
we can write by making use of the completeness of the states $\{|A\rangle\}$($\sum_A C_i^{*,A} C_{i^{\prime}}^A = \delta_{i i^{\prime}}$) : 
\begin{eqnarray}
|\langle P, T| F \rangle |^2 + \sum_{A \neq P} |\langle A, T | F \rangle |^2 & = 
& \sum_{A} |\langle A, T | F \rangle |^2 = \sum_A \left| \sum_i C_i^{*,A} C_i^P \sum_j |C_j^T|^2 t_{ij}\right|^2 \nonumber \\
& = & \sum_A \left|\sum_i C_i^{*,A} C_i^P \langle t(j) \rangle_T \right|^2 = \sum_i C_i^{*,P} C_i^P |\langle t(j) \rangle_T|^2 \nonumber \\
&= &\langle |\langle t \rangle_T|^2 \rangle_P \,\
\end{eqnarray} In a similar fashion, we can obtain the following result:

\begin{eqnarray}
|\langle P, T | F \rangle |^2 + \sum_{B \neq T} |\langle P, B | F \rangle |^2 & = & \sum_{B} |\langle P, B | F \rangle |^2 = \sum_B \left| \sum_j C_j^{*,B} C_j^T \sum_i |C_i^P|^2 t_{ij} \right|^2 \nonumber \\
& = & \sum_B \left| \sum_j C_j^{*,B} C_j^T \langle t(i) \rangle_P \right|^2 = \sum_j C_j^{*,T} C_j^T |\langle t(i) \rangle_P|^2  \nonumber \\ 
&=& \langle |\langle t \rangle_P|^2 \rangle_T \,\,,
\end{eqnarray}
and
\begin{eqnarray}
\sum_{A \neq P; B \neq T} |\langle A, B | F \rangle |^2 & = & 
\langle F | F \rangle -  |\langle P, T | F \rangle |^2 - \sum_{A \neq P} |\langle A, T | F \rangle |^2 - \sum_{B \neq T} |\langle P, B | F \rangle |^2  \nonumber \\
& = &  \langle |t|^2 \rangle_{P,T} - \langle |\langle t \rangle_T|^2 \rangle_P - \langle |\langle t \rangle_P|^2 \rangle_T  +  |\langle t \rangle_{P,T}|^2 
\end{eqnarray} Thus, based on the aforementioned relations, we can deduce the related cross-sections in the impact parameter space in the following manner:

\begin{itemize}
    \item The elastic cross-section:
\begin{eqnarray}
\frac{d^2 \sigma_{el}}{d^2b} = |\langle P, T | F \rangle |^2 =  |\langle t \rangle_{P,T}|^2 \,\,;
\label{eq:elas}
\end{eqnarray}
\item The projectile single diffractive cross-section : 
    
\begin{eqnarray}
\frac{d^2 \sigma_{sd}^P}{d^2b} = \sum_{A \neq P} |\langle A, T | F \rangle |^2 = \langle |\langle t \rangle_T|^2 \rangle_P - |\langle t \rangle_{P,T}|^2 \,\,; 
\label{eq:sdp}
\end{eqnarray}
    \item The target single diffractive cross-section :
\begin{eqnarray}
\frac{d^2 \sigma_{sd}^T}{d^2b} = \sum_{B \neq T} |\langle P, B | F \rangle |^2 = \langle |\langle t \rangle_P|^2 \rangle_T - |\langle t \rangle_{P,T}|^2 \,\,;
\label{eq:sdt}
\end{eqnarray}
\item The double diffractive cross-section:
\begin{eqnarray}
\frac{d^2 \sigma_{dd}}{d^2b} &=& 
\sum_{A \neq P; B \neq T} |\langle A, B | F \rangle |^2 \nonumber \\
& =  &  \langle |t|^2 \rangle_{P,T} - \langle |\langle t \rangle_T|^2 \rangle_P - \langle |\langle t \rangle_P|^2 \rangle_T  +  |\langle t \rangle_{P,T}|^2 \,\,.
\label{eq:sdd}
\end{eqnarray}
\item Moreover,  the total single diffractive cross section is expressed as :
\begin{eqnarray}
\frac{d^2 \sigma_{sd}}{d^2b} = \frac{d^2 \sigma_{sd}^P}{d^2b} + \frac{d^2 \sigma_{sd}^T}{d^2b} = \langle |\langle t \rangle_T|^2 \rangle_P + 
\langle |\langle t \rangle_P|^2 \rangle_T - 2 |\langle t \rangle_{P,T}|^2 \,\,,
\label{eq:sdtot}
\end{eqnarray}
and the total diffractive cross-section as :
\begin{eqnarray}
\frac{d^2 \sigma_{diff}}{d^2b} = \frac{d^2 \sigma_{sd}}{d^2b} + \frac{d^2 \sigma_{dd}}{d^2b} =  \langle |t|^2 \rangle_{P,T} - |\langle t \rangle_{P,T}|^2 \,\,.
\label{eq:diftot}
\end{eqnarray}
\item Finally, using the optical theorem, the total cross-section is given by
\begin{eqnarray}
\frac{d^2 \sigma_{tot}}{d^2b} = 2  \, \Im \left\{\langle t \rangle_{P,T} \right\} \,\,.
\label{eq:tot}
\end{eqnarray}
\end{itemize}

To compute the required average over the configurations in both the projectile and the target, necessary for obtaining these cross-sections and encompassing the entire space of parton configurations, we perform a mapping of this space onto the domain of real positive numbers. This mapping is established under the assumption that the distinct configurations $\mathbb C_i$ can be effectively represented by a continuous distribution, where each configuration is assigned a corresponding probability $P_{hi}(\mathbb C_i)$.

Accordingly, we can substitute a discrete summation with a continuous one, and this leads to the following correspondences :
\begin{align}
 \sum_i |C_i^P|^2 & \to \int d\mathbb C_1 P_{h1}(\mathbb C_1) \text{ for the projectile,}\\
 \sum_i |C_i^T|^2 & \to \int d\mathbb C_2 P_{h2}(\mathbb C_2) \text{ for the target}
\end{align}
\noindent and

\begin{equation}
 t_{ij}(b,s) \to t(b,s,\mathbb C_1,\mathbb C_2) \,\,,
\end{equation} where the different configurations are clearly displayed.

In order to reveal the role of taking into account a full parton configuration space based on this formalism, we need a model for the elastic scattering amplitude $t(b,s,\mathbb C_1,\mathbb C_2)$,  and  the probability distribution  $P_{hi}(\mathbb C_i)$, which will explicitly be presented in the following section.

\section{\label{sec:Exp mod}Explicit model and data}

The elastic hadron scattering amplitude $t(b,s,\mathbb C_1,\mathbb C_2)$ is a complex function that describes the probability of two hadrons scattering off each other at a given energy and impact parameter. At high energies, this amplitude can become very large, which violates the unitarity condition that the probability of any physical process cannot exceed unity. To restore unitarity, we can use a process called unitarization. This involves modifying the amplitude in a way that satisfies unitarity while preserving its physical properties. As has been stated in the introduction, we will consider that  $t(b,s,\mathbb C_1,\mathbb C_2)$ is given by the U-Matrix form \cite{Cudell_2009},  as the sum of all n-pomeron exchange contributions from the single-pomeron scattering amplitude which in turn is related to the expected number $\chi(b,s,\mathbb C_1,\mathbb C_2)$  of interactions between partons of the incident hadrons for a given combination of configurations $\mathbb C_1$ and $\mathbb C_2$:

\begin{equation}
	t(b,s,\mathbb C_1,\mathbb C_2) = \frac{\chi(b,s,\mathbb C_1,\mathbb C_2)}{1 - \ii \,  \chi(b,s,\mathbb C_1,\mathbb C_2)/2}
	\label{eq:umatx}
\end{equation} In order to simplify the calculation of the elastic scattering amplitude, we suppose that the expected number of interactions between
partons can be expressed as a product of the single-Pomeron scattering amplitude and some functions of impact parameter and configurations. 

\begin{equation}
\chi(b,s,\mathbb{C}_1,\mathbb{C}_2) = f(b,\mathbb{C}_1,\mathbb{C}_2) \cdot \chi_P(s, b)
\end{equation} This factorization is based on the idea that the configurations dependence of $\chi(b,s,\mathbb C_1,\mathbb C_2)$ can be separated from the energy dependence, which is described by the single-Pomeron scattering amplitude. This assumption is based on the fact that the energy dependence of the elastic scattering amplitude is dominated by the exchange of a single Regge pole, the pomeron, which is independent of the specific hadronic configurations involved in the scattering process

In addition, if we assume that the distribution of parton configurations is independent of the impact parameter, which means that the parton density inside the hadron is the same at all points in space and that the hadron can be treated as a collection of independent partons, then  we can write
\begin{equation}
     \chi(b,s,\mathbb C_1,\mathbb C_2) = 
      \chi_P(s, b) \cdot \alpha(\mathbb C_1)\alpha(\mathbb C_2)
\end{equation} where the functions $\alpha(\mathbb C_i)$  depend on the configurations of the incident hadrons. Therefore, we have
%

\begin{equation}
 \int d\mathbb C_1 \int d\mathbb C_2 P_{h1}(\mathbb C_1) P_{h2}(\mathbb C_2) t(b, s, \mathbb C_1,\mathbb C_2) =  \int_0^\infty d\alpha_1\int_0^\infty d\alpha_2 p(\alpha_1)p(\alpha_2) t(b, s, \alpha_1, \alpha_2)
\end{equation} where the functions $p(\alpha_i)$ are defined by
\begin{equation}
 p(\alpha_i) = \int d\mathbb C_i P_{hi}(\mathbb C_i)\delta[\alpha(\mathbb C_i) - \alpha_i],
\end{equation}
which satisfy the following constraints:
\begin{equation}
 \int_0^\infty d\alpha_i \, p(\alpha_i) = 1,
 \label{eq:norm_p_alpha}
\end{equation}
and 
\begin{equation}
 \int_0^\infty d\alpha_i \, \alpha_i \, p(\alpha_i)   = 1 \quad (i=1,\,2) \,\,.
 \label{eq:average_p_alpha}
\end{equation}

Accordingly, we can implicitly take into account of an infinite number of inelastic channels by using the function of a real positive variable, the probability distribution $p(\alpha)$ representing the fluctuations of the hadron configurations with some extension defined by its variance. This generalizes the GW approach to a multichannel framework, as demonstrated in \cite{Lipari2009}, where the connection between the discrete and continuous multi-channel GW was established. 

Thus, The averaging over the configurations appearing in Eqs. (\ref{eq:elas}) -- (\ref{eq:tot}) will be determined as follows:

\begin{itemize}
 \item Mean value computed over the configurations of the projectile: 
 \begin{equation}
 \langle t^n \rangle_P = \int_0^\infty d\alpha_1 \, p(\alpha_1) \, t^n(b,s,\alpha_1,\alpha_2);
 \label{eq:average_proj}
 \end{equation}
 
\item Mean value computed over the configurations of the target : 
\begin{equation}
\langle t^n \rangle_T = \int_0^\infty d\alpha_2 \, p(\alpha_2) \, t^n(b,s,\alpha_1,\alpha_2);
\label{eq:average_tar}
\end{equation}
 
\item Mean value computed over the configurations of the projectile and the target :  
 \begin{eqnarray}
 \langle t^n \rangle_{PT} = \int_0^\infty d\alpha_1 \int_0^\infty d\alpha_2 \, p(\alpha_1) \, p(\alpha_2) \, t^n(b,s,\alpha_1,\alpha_2)
 \label{eq:average_proj_tar}
 \end{eqnarray}
\end{itemize}

An advantage to the method disclosed here is that it considers the entirety of the parton configuration space. Nevertheless, it should be noted that the probability distribution $p(\alpha_i)$, remains unknown. We do anticipate, however, that this distribution will exhibit the following characteristics: it needs to be defined for positive values of its variable $\alpha$ and  have the predicted limit, $p(\alpha)\to\delta(\alpha-1)$, when its variance reaches zero, which is equivalent to no fluctuations and satisfies the above constraints \ref{eq:norm_p_alpha} and \ref{eq:average_p_alpha}. In order to satisfy these properties,  we use for the probability distribution $p(\alpha_i)$, the gamma distribution, with variance $w$,
  
\begin{equation}
  p(\alpha_i) = \frac{1}{w\Gamma(1/w)}\left(\frac{\alpha_i}{w}\right)^{-1+1/w}e^{-\alpha_i/w}\,\, 
  \label{eq:p_alpha}
\end{equation} Since we are accounting for the collision of identical hadrons, we will  suppose that the variance $w$ of the distribution is independent of $i$. This assumption enables us to compute the average over configurations Eqs.~\eqref{eq:average_proj}, \eqref{eq:average_tar} and \eqref{eq:average_proj_tar}  needed to determine the various observables Eqs.~\eqref{eq:elas}-~\eqref{eq:tot}.

To complete the description of our model, we parameterize the single-pomeron scattering amplitude, as the {\it Ansatz} put forth in \cite{Vanthieghem_2021}  for comparison purposes :



\begin{equation}
	a_{P}(s,t) =g_{pp}^2\, \fform_{pp}(t)^2 \left( \frac{s}{s_0} \right)^{\alpha(t)}\, \xi(t),
	\label{eq:amp_t}
\end{equation} where   $\alpha(t)$ is the pomeron  trajectory,  $\fform_{pp}(t)$ is the proton elastic form factor, and $g_{pp}$ is  the coupling pomeron-proton-proton, with $\xi(t)$ the signature factor

\begin{equation}
	\xi(t) =-e^{-i\pi\alpha(t)\over 2},
\label{eq:amp_t2}
\end{equation} where a full complex rather than a purely imaginary one was chosen in order to meet the elastic amplitude's analyticity constraint, which is essential to respecting causality. Regarding the proton elastic form factor, although the exact functional form is not very important as we want to make a comparison with the two-channel model, we shall consider here a dipole form factor :

\begin{equation}
\fform_{pp}={1\over (1-t/t_{pp})^2}
\label{form}
\end{equation}
Using an exponential form factor \{$F_1 = \exp \left( R_0 t \right)$\},
instead of the dipole form, leads to slightly poorer fits \cite{Bhattacharya_2021}.
The pomeron trajectory is close to a straight line \cite{Cudell:2003dz}  and we take it to be

\begin{equation}
	\alpha(t)=1+\epsilon+\apr t.
	\label{eq:amp_t3}
\end{equation} In the impact-parameter space representation, where  the Fourier transform of the amplitude $a_P(s, t)$ rescaled by $2s$ is equivalent to a partial wave, we have : 

\begin{equation}
	\chi_P(s, \bvec{b}) = \int \frac{\ud^2\bvec{q}}{\left( 2\pi \right)^2}
	\frac{a_P(s,t)}{2s} \ee^{\ii \bvec{q}\cdot \bvec{b}}.
	\label{eq:Gsb}
\end{equation} and by the unitarisation procedure we map the amplitude $\chi_P(s, b)$ to the physical amplitude $t(s,b)$, which in turn bears the same relation as Eq.~\eqref{eq:Gsb}, but this time to the physical amplitude : 

\begin{equation}
	t(s, \bvec{b}) = \int \frac{\ud^2\bvec{q}}{\left( 2\pi \right)^2}
	\frac{A(s,t)}{2s} \ee^{\ii \bvec{q}\cdot \bvec{b}}.
	\label{eq:tsb}
\end{equation} Thus, using the assumptions made in this model, we can make specific predictions and conclusions about the hadronic collisions at high energy and, at the same time, test the hypotheses that were adopted. 

Before presenting our results in the subsequent section, we list here the model parameters that will be set by the data fit as well as the experimental data employed. The model parameters are the following:  $\epsilon$ and $\alpha'$, which are associated with the Pomeron trajectory, as well as $g_{pp}$ and $t_{pp}$ linked to the proton-pomeron $p I\!\!P p$ vertex, and the variance $\omega$ of the probability distribution.
We employ experimental data above 100 GeV as  we are concerned with  high energy effects induced in  $ p\porpb $ cross sections. And as we aim at looking into the impact of putting in place a multi-channel model in order to describe hadronic interactions and comparing it with the two-channel one, the same data set \footnote{see compilation in \cite{Vanthieghem_2021}} as in \cite{Vanthieghem_2021}, which involves statistical as well as systematic errors and combines them in quadrature, is used. The fitting process was conducted using the class Minuit2 from ROOT \cite{Hatlo_2005} and the MIGRAD algorithm. The fitting was performed by minimizing the $\chi^2$ value, and the uncertainties in the free parameters were calculated with a $1 \sigma$ confidence level, which was used to determine the error band.

\section{\label{sec:result}Results and discussion}

\begin{figure}[!htpb]
\centering
\subfloat{\includegraphics[height=0.4\textwidth]{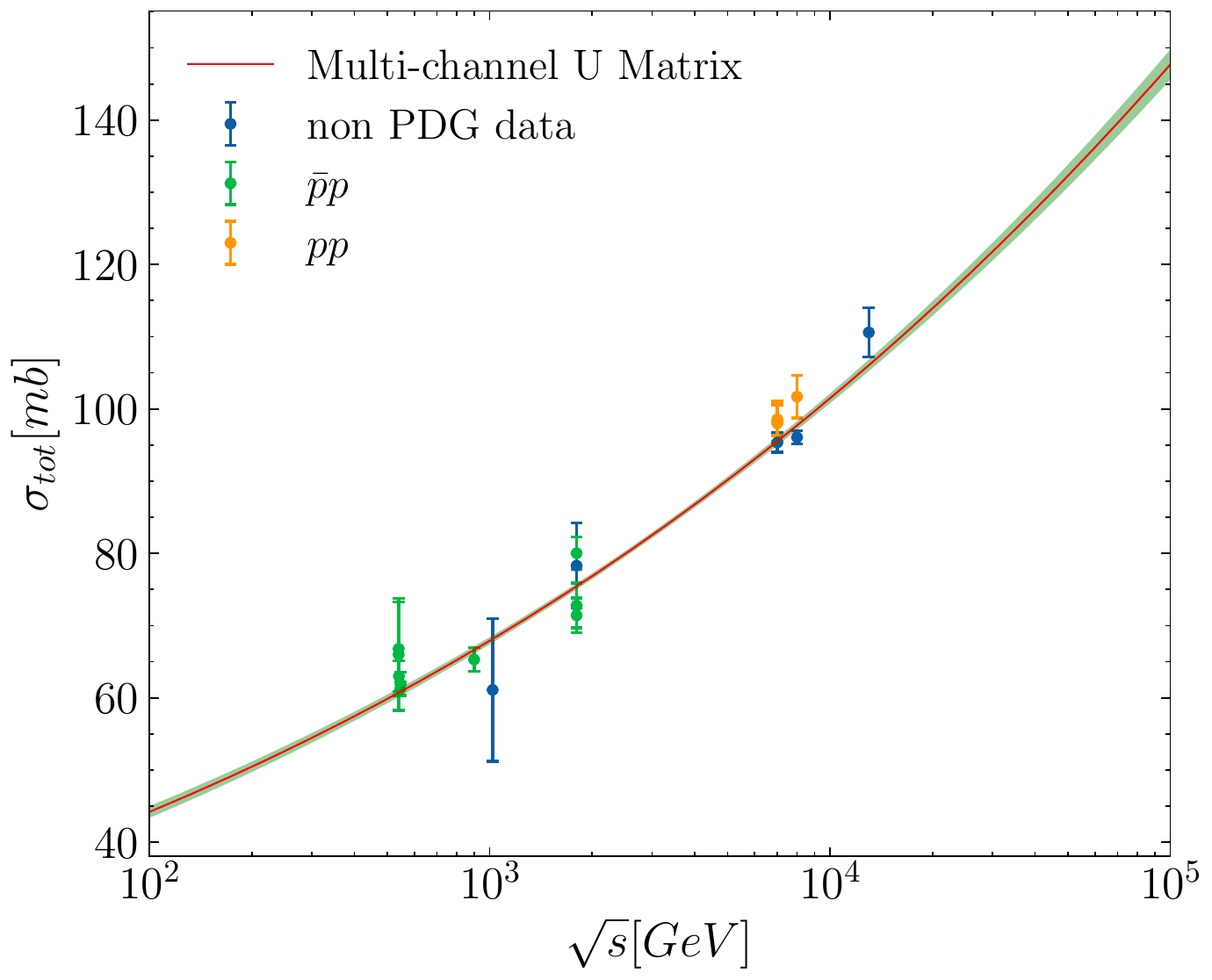}}
\subfloat{\includegraphics[height=0.4\textwidth]{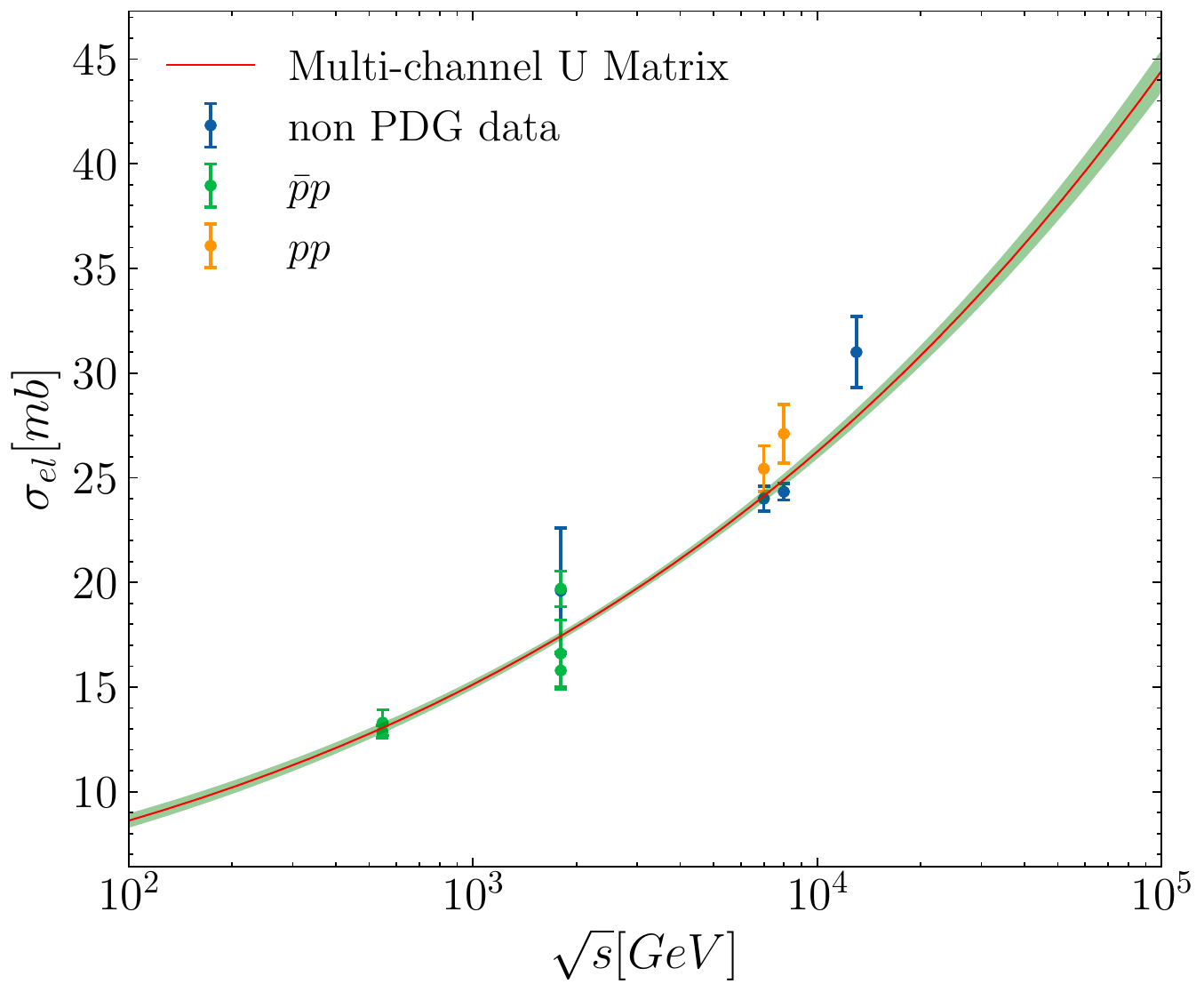}} \\
\subfloat{\includegraphics[height=0.4\textwidth]{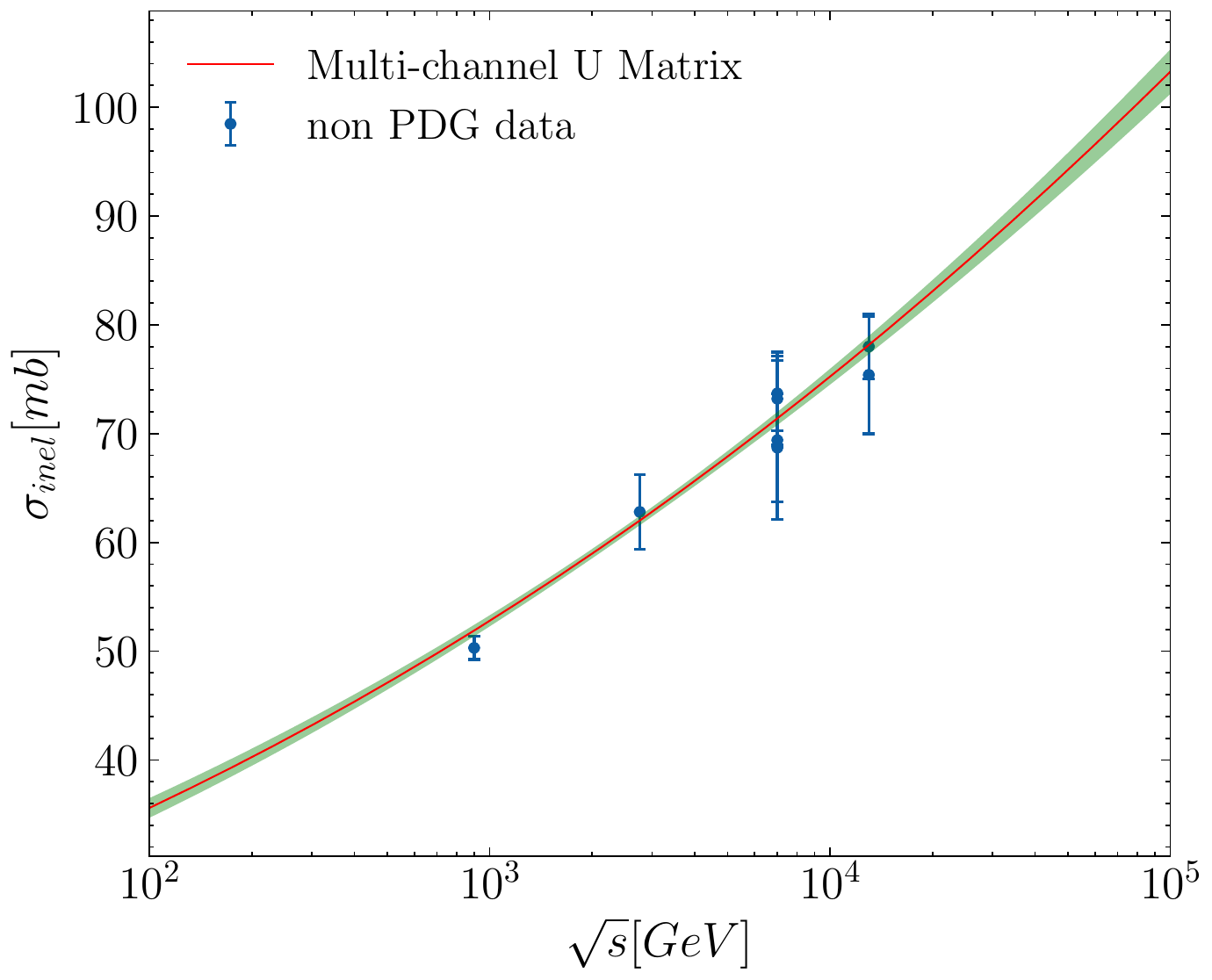}}
\subfloat{\includegraphics[height=0.4\textwidth]{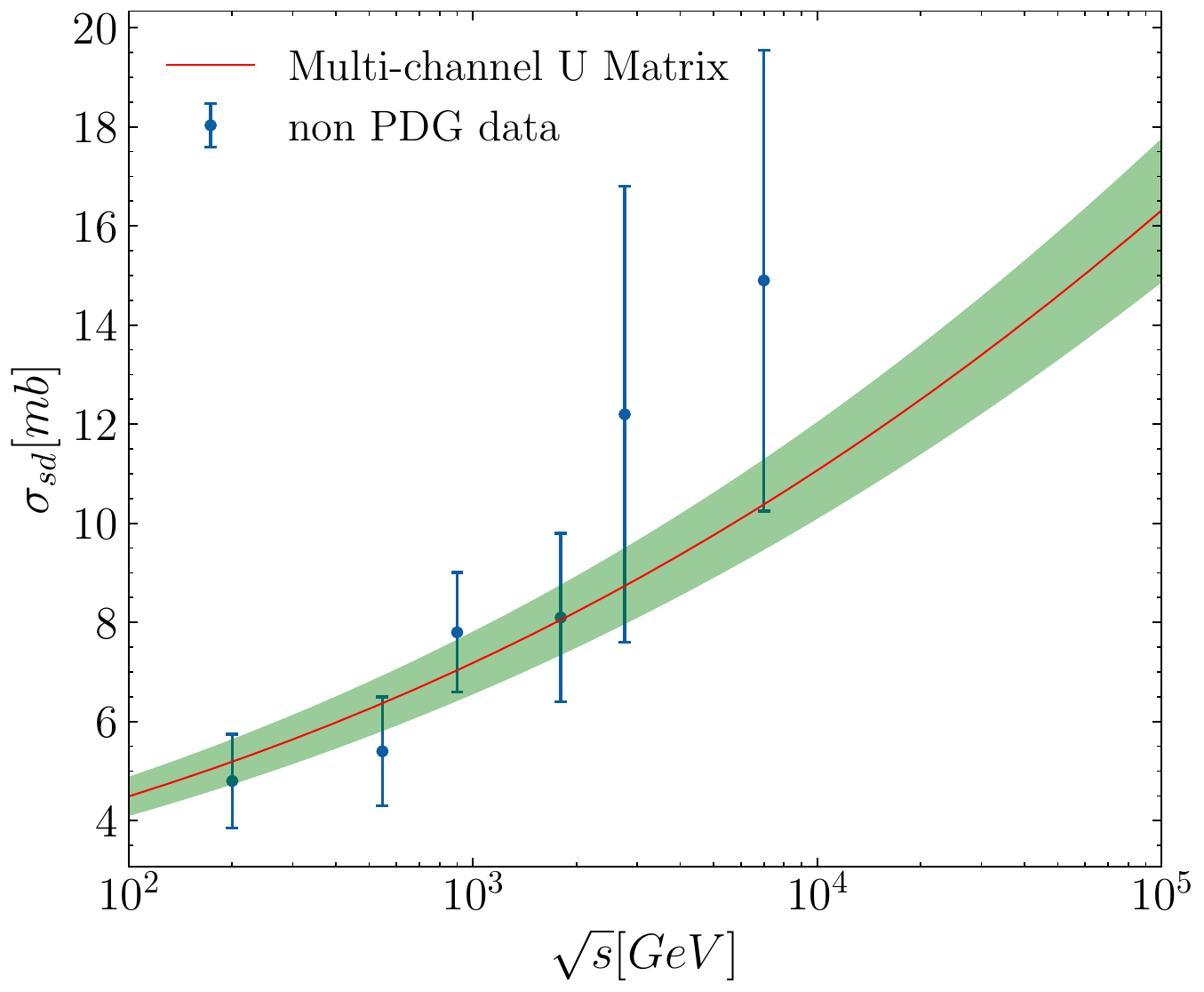}}
\caption{\label{fig:bfxsec} Total, elastic, inelastic
and single diffractive cross sections with the Multi-channel model and the $1\sigma$ error bands around the fitted curve obtained with best-fit  parameters.}
\end{figure}

The results of our multi-channel model  are provided in Fig.~\ref{fig:bfxsec} and Table \ref{tab:bfits_param} using the formalism previously outlined. As can be seen from these findings, the multi-channel model describes well the total, elastic, inelastic, and single-diffractive cross-sections, with a $\chi^2  $ /d.o.f of 1.328.  These outcomes are actually in line with those obtained using the $U$-matrix two-channel model with a $\chi^2  $ /d.o.f of 1.316 \cite{Vanthieghem_2021}, which shows a difference of only 0.012 in the $\chi^2  $ /d.o.f.

\begin{figure}[!htpb]
\subfloat{\includegraphics[height=0.4\textwidth]{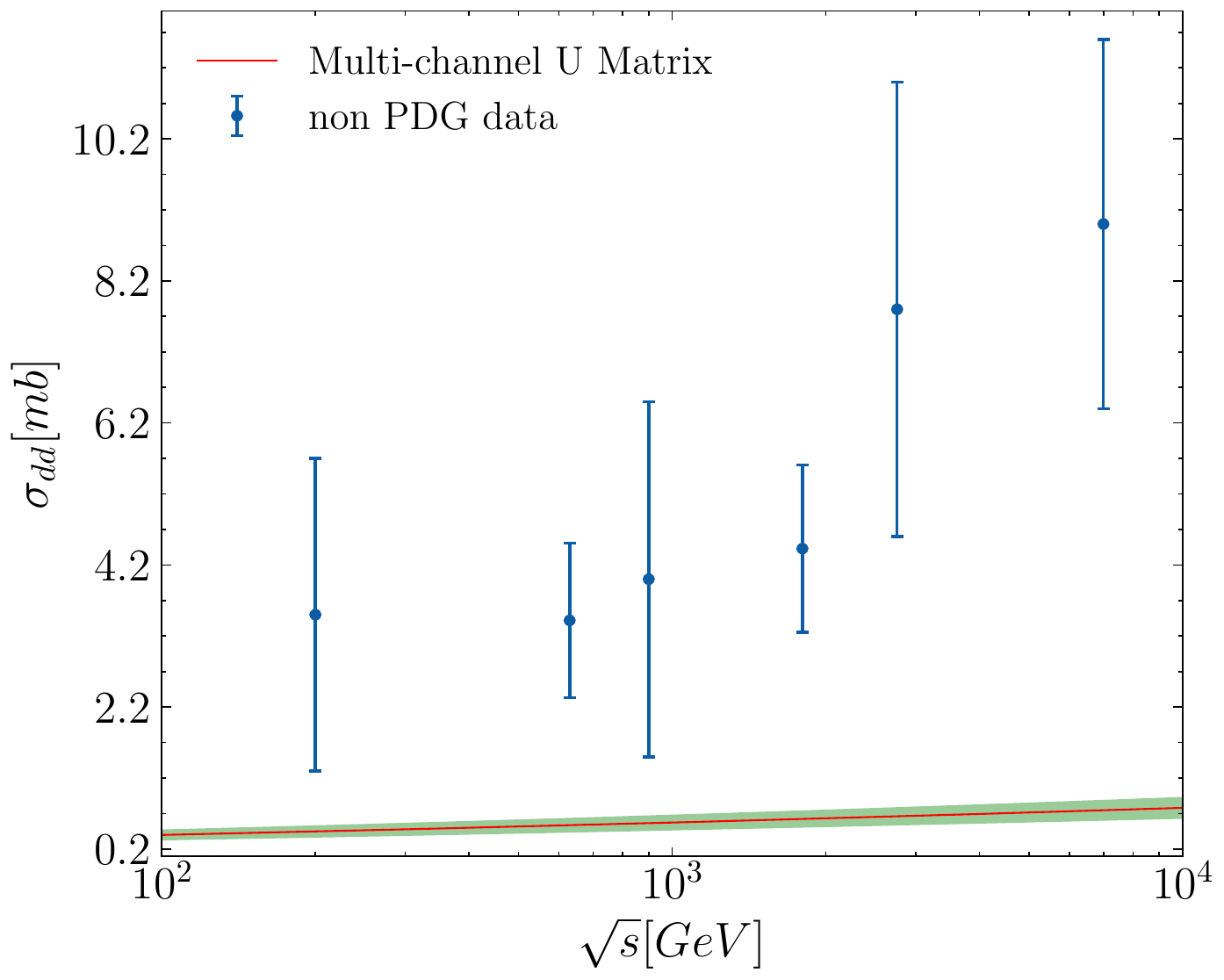}}
\subfloat{\includegraphics[height=0.4\textwidth]{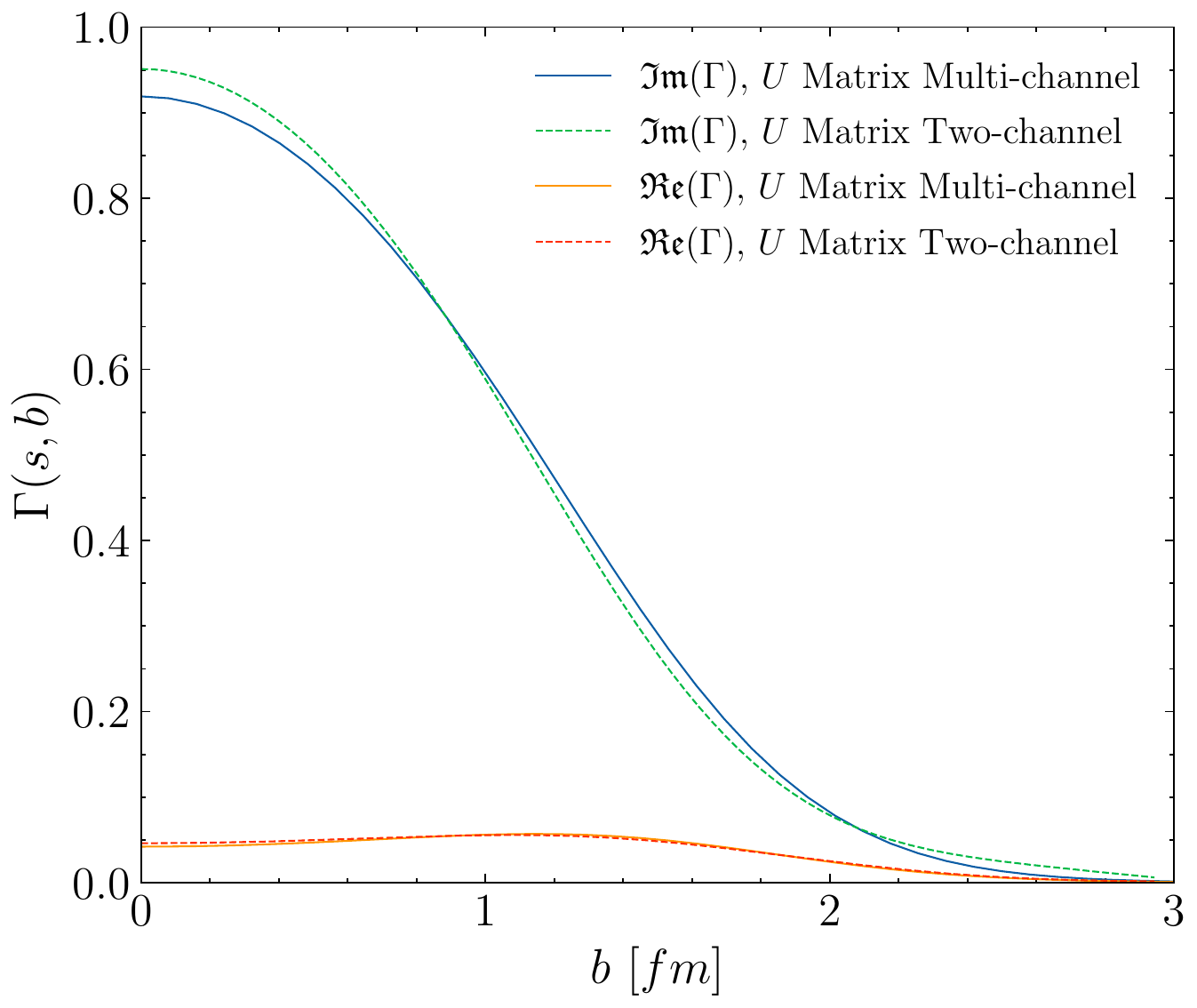}}
\caption{\label{fig:dd_xsec_elastic_prof} The double diffractive cross section with $1\sigma$ error bands around the predicted curve obtained with the Multi-channel model (left panel). The 
real and imaginary parts of the elastic profile function $\Gamma(s,b)$ at $\sqrt{s}=13$ TeV with the U Matrix scheme for the two and multi-channel cases (right panel).}
\end{figure}

The difference between the two $\chi^2  $ obtained in both models is marginal. As a matter of fact, it is somewhat surprising that there is no improvement with the multi-channel model given that the latter was expected to describe  the diffractive phenomenon, in particular,  better than the two-channel one. The reason for this similarity between the two models can be attributed to the unitarization process employed in both cases. Specifically, both models adhere to the same unitarity constraint, known as the $U$-matrix scheme. This can be observed in Fig.~\ref{fig:dd_xsec_elastic_prof} (right panel), where the impact-parameter space representation showcases that the imaginary and real components of the elastic profile function at a specific energy, such as 13 TeV, are nearly identical in both models. Furthermore, it is evident that these components do not surpass the black disk limit, indicating consistency with the principles of unitarity. Most importantly, based on the similarity in the obtained $\chisq/d.o.f.$ values between the $U$-matrix multi-channel model and the eikonal two-channel one \cite{Vanthieghem_2021}, it seems likely that the factorization assumption adopted in the former case may not be applicable. Specifically, if there is a correlation between the pomeron exchanges, then the impact parameter and configuration dependence of the
scattering amplitude may not be separable from the energy dependence carried by the single-pomeron exchange. In this case, the distribution of parton configurations may depend on the impact parameter, and the average number of interactions at a fixed impact parameter and center of mass energy may not be separable from the configuration dependence of the incident hadrons.

\begin{table*}[!htpb]
\begin{adjustbox}{width=\columnwidth,center}

\begin{tabular}{|c||c|c|c|c|c|c|}
   \hline
  Model   & $\epsilon $
            & $\apr $ (GeV$^{-2}$)
            & $ g_{pp} $
            & $ t_{pp} $ (GeV$^2$)
            & $ \omega $ 
            & $ \chisq/\dof $ \\
            \hline 
   Multi-channel  & $ 0.11 \pm 0.003 $
            & $ 0.29 \pm 0.04 $
            & $ 8.25 \pm 0.2 $
            & $ 2.06 \pm 0.75$
            & $ 0.59 \pm 0.06$
            & $ 1.328 $ \\
   \hline
\end{tabular}

\end{adjustbox}
\caption{\label{tab:bfits_param}\chisq/\dof\ and best-fit parameters obtained with the $U$-matrix Multi-channel model.}
\end{table*}
Double diffractive cross-section measurements are not included in our fits and our prediction based on the model presented in this study doesn't reproduce these data in spite of considering an infinite parton configurations space, which corroborates the result reported in the context of a two-channel model \cite{Vanthieghem_2021}, as illustrated in Fig.~\ref{fig:dd_xsec_elastic_prof} (left panel).  In fact, a proper description of this cross-section requires the introduction of an additional contribution due to the
Pomeron-enhanced diagrams involving Pomeron-Pomeron interactions which is non GW. As the energy increases, more diagrams illustrating complicated topologies become involved. Consequently, the consistent treatment of these enhanced corrections proves to be a very challenging task \cite{Ostapchenko:2006vr}.

\begin{figure}[!htpb]
\centering
\subfloat{\includegraphics[width=0.5\textwidth]{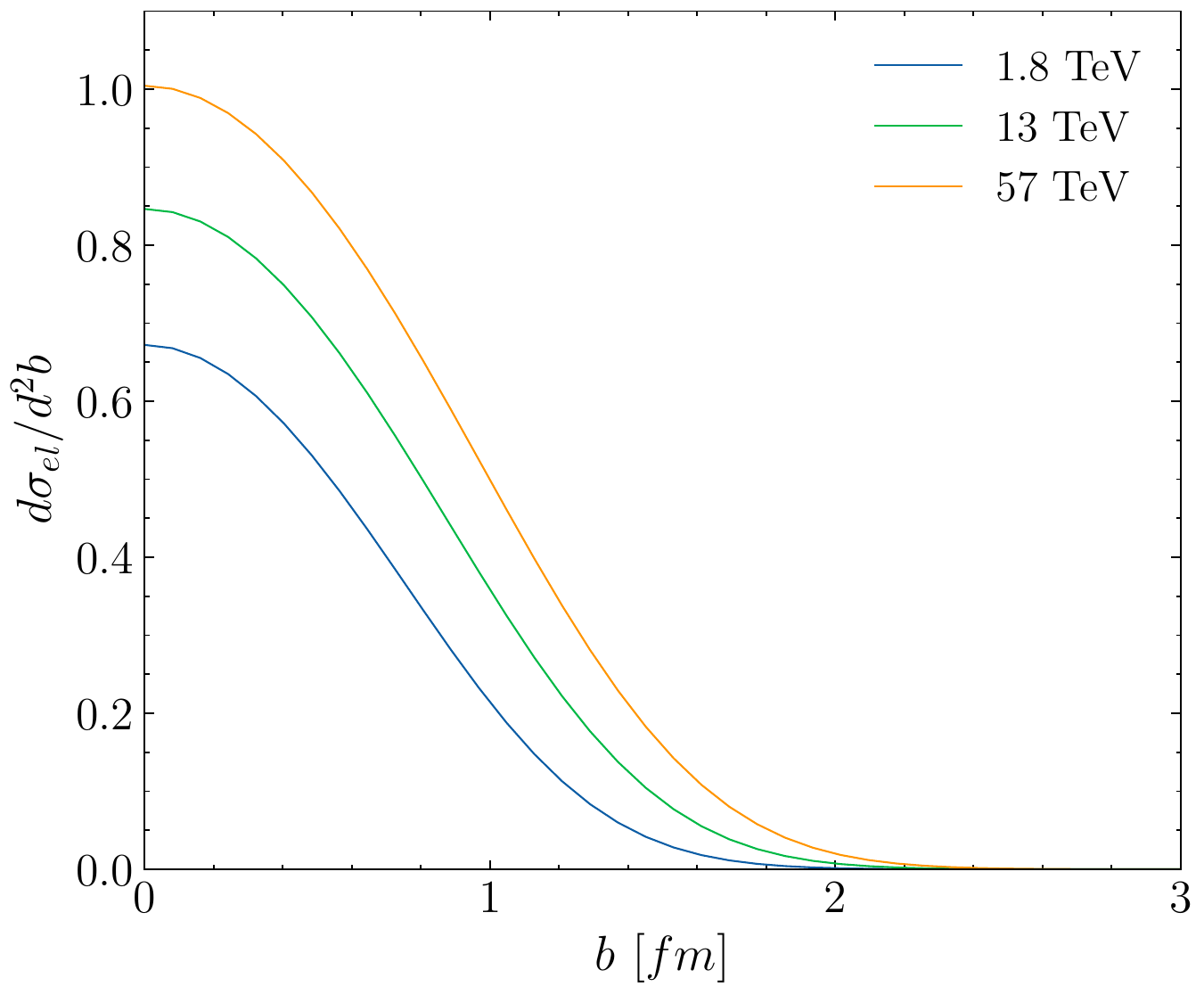}}
\subfloat{\includegraphics[width=0.5\textwidth]{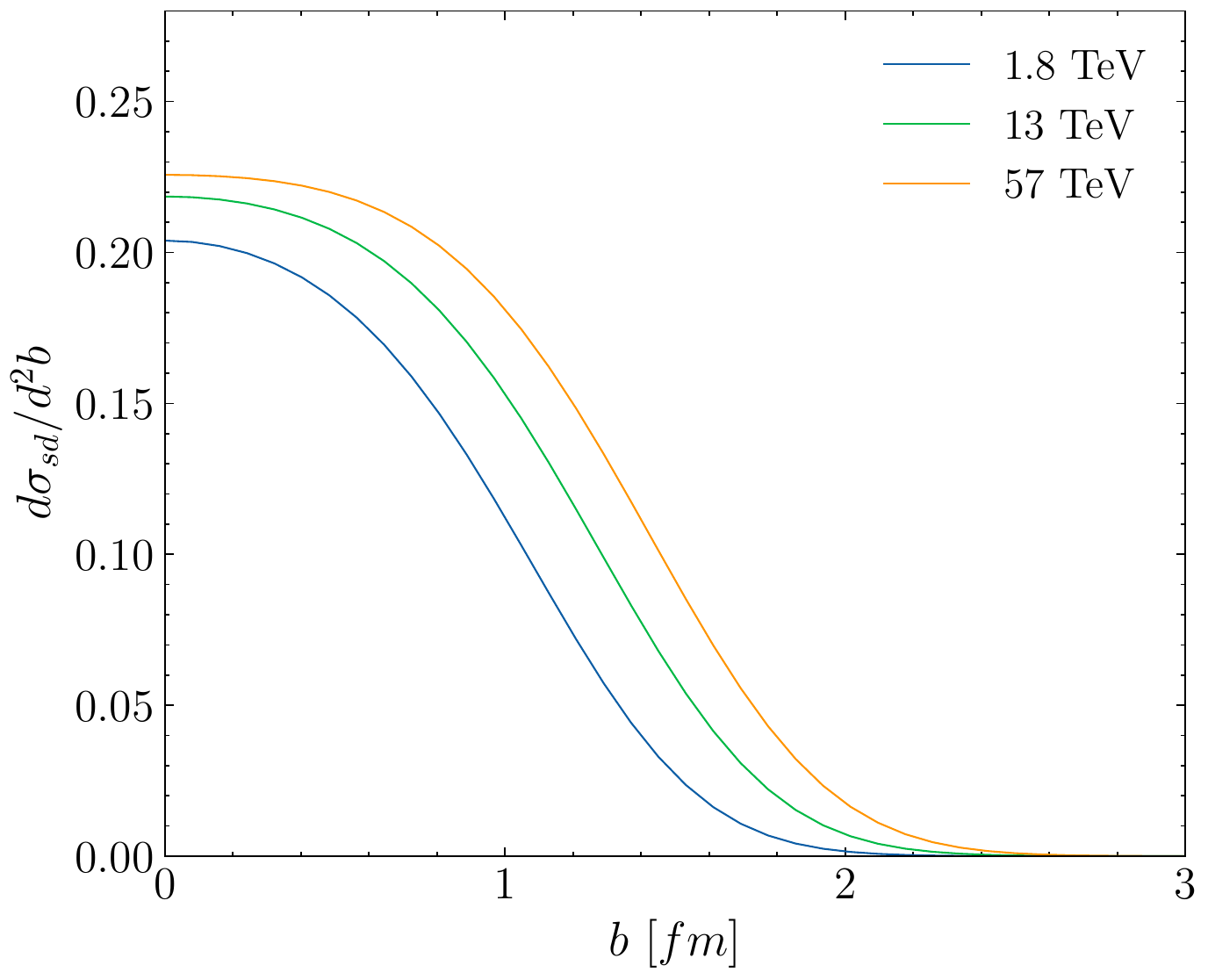}}\\ 
\subfloat{\includegraphics[width=0.5\textwidth]{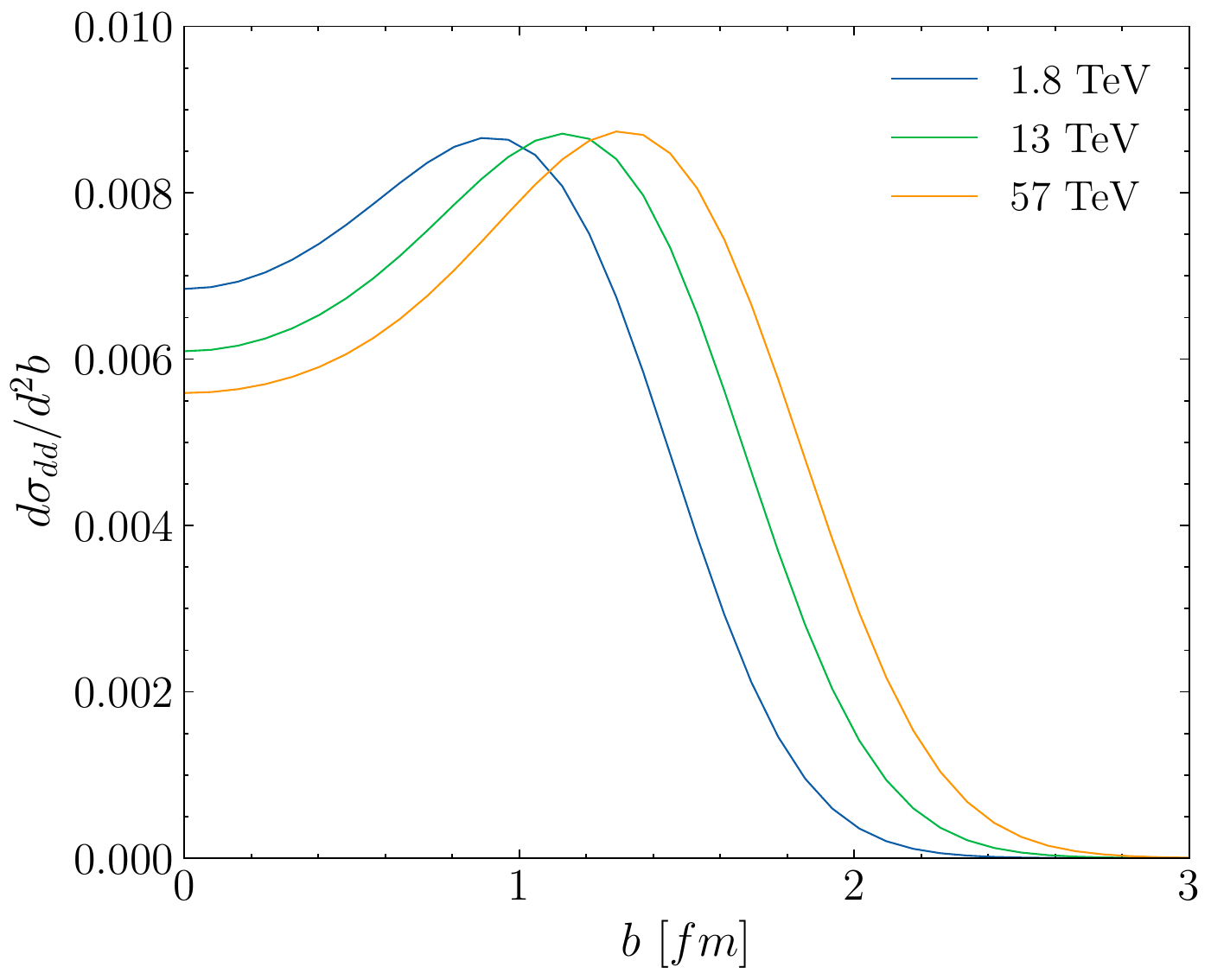}}
\caption{\label{fig:dxsec_b} Multi-channel model predictions for the energy dependence of the elastic, single diffractive, and double diffractive  differential cross sections in impact parameter space.}
\end{figure}

Fig.~\ref{fig:dxsec_b} displays the predictions for the energy evolution of the cross sections in the impact parameter space from Tevatron to cosmic ray energies. The elastic, single-diffractive, and double-diffractive differential cross sections are shown in the top right, bottom left and right panels, respectively. It can be seen that the elastic scattering is primarily central and increases with energy. This result is comparable to the one reported in \cite{Broilo2020}. In contrast, it gets much closer to the black disk limit at cosmic ray energies. The behaviour of the single diffractive differential cross-section is similar to that of the elastic scattering. At b = 0, it is mostly central and has a magnitude that grows with energy, but it is smaller than the one of the elastic scattering. It also declines more slowly than the elastic cross-section as b rises. Similar behaviours of the unintegrated profile for the single diffractive cross section at low mass are predicted by the Kolevatov and Boreskov model presented in \cite{Kolevatov:2012vu}. This result contrasts with the one presented in \cite{Broilo2020}, where the total single diffractive cross section becomes more peripheral, with a maximum  moving to a higher impact parameter as the energy rises. Furthermore, as the c.m. energy increases, the magnitude of the SD cross-section at b = 0 decreases. This rather contradictory result might be attributed to the use of two different unitarization schemes. The double diffractive cross-section becomes more peripheral when energy rises. Nevertheless, its magnitude at b = 0 diminishes as c.m. energy increases. This result is in line with that obtained in \cite{Broilo2020}. A note of caution with regard to the shape of the unintegrated profile for the double diffractive cross section is due here since it is not well described.

\begin{figure}[!htpb]
\centering
\subfloat{\includegraphics[width=0.5\textwidth]{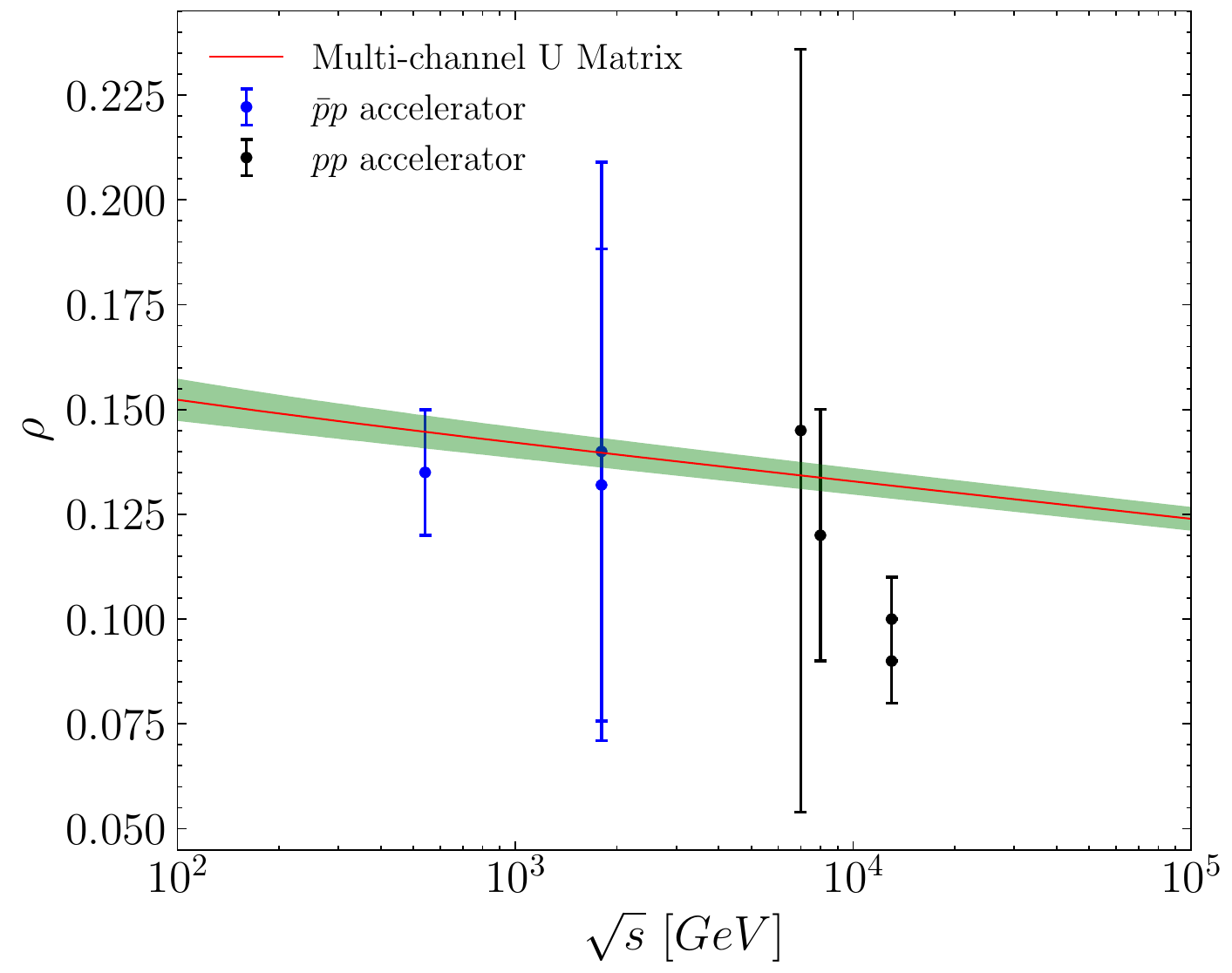}} 
\subfloat{\includegraphics[width=0.5\textwidth]{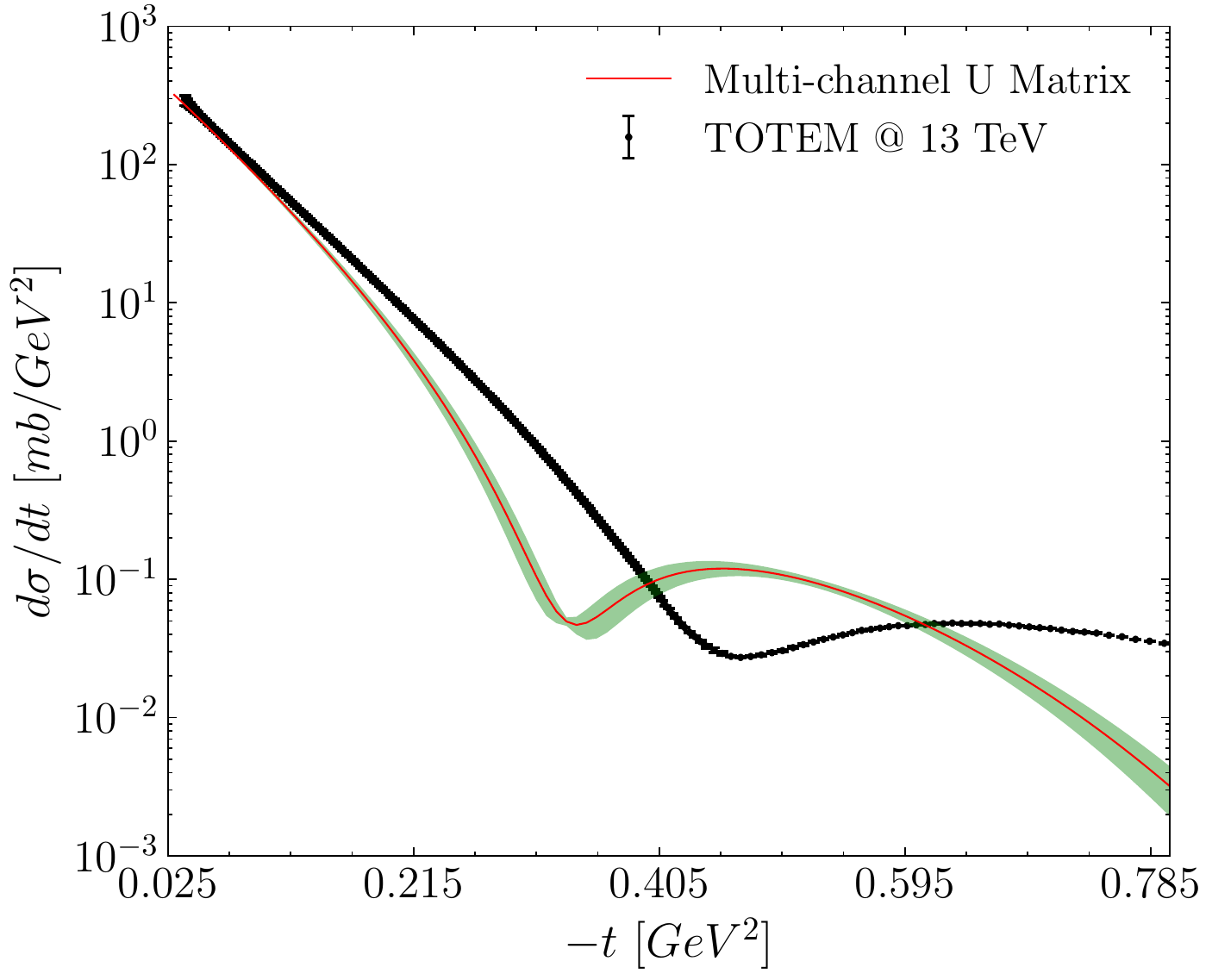}} 
\caption{\label{fig:rho_dsdt_13tev} Multi-channel model predictions for the $\rho$ parameter (left panel) and for the elastic differential cross-section at 13 TeV (right panel).}
\end{figure}

The $\rho$ parameter, i.e., the ratio of the real part of the elastic scattering amplitude to its imaginary part has been studied in several experiments at different centre-of-mass energies. In spite of the fact that $\rho$ data are not used in our fits, we can estimate its values at various $\sqrt{s}$ by using our best-fit parameters and then compare these predictions with the experimental data. Fig.~\ref{fig:rho_dsdt_13tev} (left panel) illustrates our predicted values for this observable.  As can be seen from this figure, while our model furnishes a reasonable description for this parameter at various high energies, it is unable to estimate the TOTEM data at 13 TeV since the $1\sigma$ error band of the model doesn’t  even  reach the error bars of these data.

This finding can be explained by the fact that an odderon contribution, which emerged from the TOTEM and D0 experiments is required to be included, implying distinct energy dependencies of the $pp$ and $p\bar{p}$ cross sections \cite{PhysRevLett.127.062003}.

Fig.~\ref{fig:rho_dsdt_13tev} (right panel) shows our prediction for the elastic differential cross-section  in function of the transverse momentum  in the context of a $pp$ collision at 13 TeV.   It is evident from the figure that while the model describes the experimental data for the elastic differential cross-section at small values of squared momentum transfer $q^2$, neither the position of the dip nor the behaviour at large $q^2$ is adequately described.  Similar outcomes have already been reported in a number of previous studies \cite{Broilo2020, Flensburg:2008ag}, which stresses the need for an improvement of the present model. Most importantly, this result points out that taking into account an entire parton configuration space doesn’t have any impact on the description of the 
elastic differential cross section, as has been found in the two-channel case \cite{KMR_PRD_2018}. Our model can be enhanced by considering a complex hadron overlap function rather than a simple dipole form factor, which is a reasonable approximation, since it is known that the 
elastic differential cross-section  depends on the description of the overlap function and, thus, on the internal structure of the incident hadrons \cite{KMR_PRD_2018}. As regards the position of the dip, it has been reported in \cite{KMR_PRD_2018} that an Odderon contribution is required.

\begin{figure}[htb]
\centering
\subfloat{\includegraphics[height=0.5\textwidth]{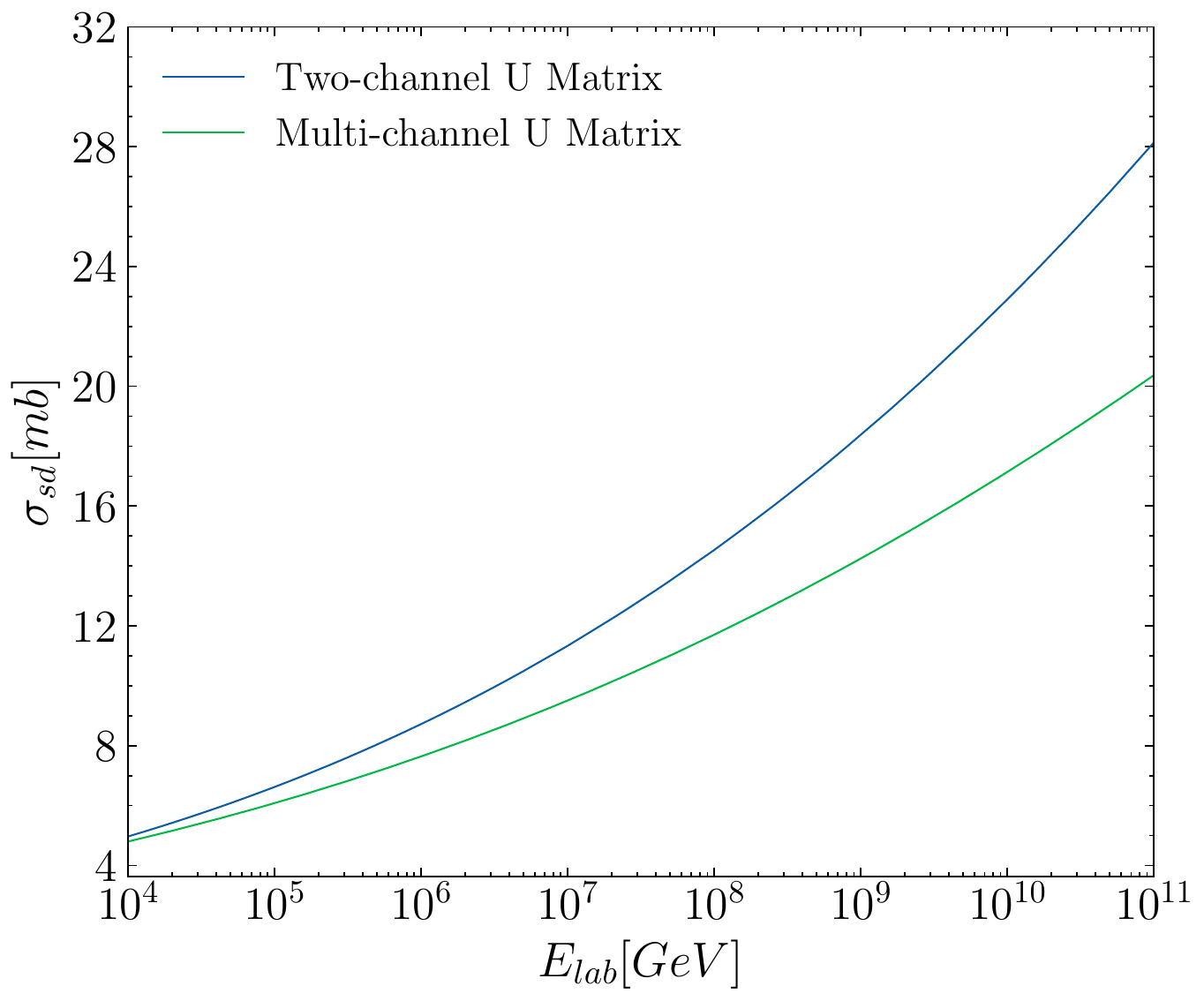}}
\caption{\label{fig:sd_xsec_elab} The growth of the single-diffractive cross-section with \emph{lab} energies up to $\sqrt{s}  = 10^{11}$ GeV for both the two and multi-channel models.}
\end{figure} 
Although the two-channel and the multi-channel models are similar in describing the various hadronic observables in comparison to the currently available data, they provide differing predictions for the single-diffractive cross-section, in particular, at ultra-high energies, as shown in the right panel of Fig.~\ref{fig:sd_xsec_elab}, where the two-channel model exhibits a faster increase with energies than the multi-channel one, using the U matrix scheme.  

This discrepancy may stem from the assumption of equal width for fluctuations in the parton configurations of colliding hadrons across different channels. This assumption simplifies the model by assuming uniform fluctuations across processes. However, in reality, fluctuations may vary between channels. To gain a deeper understanding, it would be valuable to investigate the potential for channel-dependent fluctuations. This would entail allowing the width of fluctuations to vary for each channel, thereby encompassing the unique characteristics and dynamics of individual scattering processes.

\section{\label{sec:conclusions}Conclusions}
The chief purpose of the study was to provide a phenomenological description of the hadronic interaction at high energy through extending the two-channel model into a multi-channel one using the U-matrix unitarization scheme of the elastic amplitude. 
It has been found that the multi-channel model accurately describes the total, elastic, inelastic, and single-diffractive cross-sections, with only a minor difference from the two-channel one.

In addition, in spite of considering an entire parton configuration space, the present model was not able to estimate the double-diffractive cross-section, which is in line with the results obtained with the two-channel model. In fact, a proper description of this cross-section requires the introduction of an additional contribution, i.e., pomeron interactions.

Moreover, the behaviour of the energy evolution of the various profile functions in the impact parameter space from Tevatron to cosmic ray energies was analysed.

The study has also found that the present model describes well the $\rho$ parameter at different high energies, but it is unable to estimate the TOTEM data at 13 TeV. It has been suggested that an Odderon contribution is needed to be included in order to remedy this shortcoming.

Furthermore, the elastic differential cross-section at 13 TeV was predicted. It has been shown that the model describes the experimental data for this observable at small values of squared momentum transfer $q^2$, but it doesn’t describe the position of the dip or the behaviour at large $q^2$. In this regard, it has been proposed that considering  a complex hadron overlap function instead of a simple dipole form factor as well as an odderon contribution would be  possible approaches to address this flaw. 

Last but not least, despite similarities in the way the two models describe various hadronic observables, they provide distinct predictions for the single-diffractive cross-section, especially at ultra-high energies, which represents an interesting direction for future research on ultra-high energy cosmic rays.

The paper concludes by arguing that the $U$-matrix scheme is more likely to accounting for potential correlations between pomeron exchanges. Additionally, it suggests that the two-channel model, as opposed to a multi-channel one, is adequate for modeling high-energy hadronic interactions, particularly single diffractive scattering, using the $U$-matrix scheme, even at ultra-high energies, provided that any potential pomeron correlations are disregarded.

In summary, the multi-channel model used in the paper has limitations because the probability distribution for hadron configurations is not unique, and that the impact parameter and configuration dependence of the scattering amplitude may not be separable from the energy dependence carried by the single-pomeron exchange. This would complicate the calculation of the total elastic scattering amplitude and may require a more advanced theoretical framework, which is beyond the scope of this work. 

On the whole, the findings of this study can serve as a base for future improvements of the hadronic interaction models used in cosmic ray air shower simulations.  

\acknowledgments{RO would like to thank Prof. Jean-René Cudell for his invaluable comments and fruitful discussions. Special thanks go to the computational resource provided by Consortium des Équipements de Calcul Intensif (CÉCI), funded by the Fonds de la Recherche Scientifique de Belgique (F.R.S.-FNRS) where a part of the computational work was carried out.}

\bibliographystyle{JHEP}
\bibliography{refs}

\providecommand{\href}[2]{#2}\begingroup\raggedright\begin{thebibliography}{10}

\bibitem{DENTERRIA201198}
D.~d’Enterria, R.~Engel, T.~Pierog, S.~Ostapchenko and K.~Werner,
  \emph{Constraints from the first lhc data on hadronic event generators for
  ultra-high energy cosmic-ray physics},
  \href{https://doi.org/https://doi.org/10.1016/j.astropartphys.2011.05.002}{\emph{Astroparticle
  Physics} {\bfseries 35} (2011) 98}.

\bibitem{Vanthieghem_2021}
A.~Vanthieghem, A.~Bhattacharya, R.~Oueslati and J.-R.~Cudell,
  \emph{Unitarisation dependence of diffractive scattering in light of
  high-energy collider data},
  \href{https://doi.org/10.1007/jhep09(2021)005}{\emph{Journal of High Energy
  Physics} {\bfseries 2021} (2021) }.

\bibitem{Broilo2020}
M.~Broilo, V.P.~Gon\ifmmode~\mbox{\c{c}}\else \c{c}\fi{}alves and
  P.V.R.G.~Silva, \emph{Model of diffractive excitation in $pp$ collisions at
  high energies},
  \href{https://doi.org/10.1103/PhysRevD.101.074034}{\emph{Phys. Rev. D}
  {\bfseries 101} (2020) 074034}.

\bibitem{Treleani_2008}
D.~Treleani, \emph{A multi-channel poissonian model for multi-parton
  scatterings},  2008.
\newblock 10.48550/ARXIV.0808.2656.

\bibitem{Durand_1988}
L.~Durand and H.~Pi, \emph{High-energy nucleon-nucleus scattering and
  cosmic-ray cross sections},
  \href{https://doi.org/10.1103/PhysRevD.38.78}{\emph{Phys. Rev. D} {\bfseries
  38} (1988) 78}.

\bibitem{Martynov_2020}
E.~Martynov and G.~Tersimonov, \emph{Multigap diffraction cross sections:
  Problems in eikonal methods for the pomeron unitarization},
  \href{https://doi.org/10.1103/PhysRevD.101.114003}{\emph{Phys. Rev. D}
  {\bfseries 101} (2020) 114003}.

\bibitem{Troshin:2003wu}
S.M.~Troshin and N.E.~Tyurin, \emph{{Unitarity at the LHC energies}},
  {\emph{Phys. Part. Nucl.} {\bfseries 35} (2004) 555}
  [\href{https://arxiv.org/abs/hep-ph/0308027}{{\ttfamily hep-ph/0308027}}].

\bibitem{Troshin_2020}
S.M.~Troshin and N.E.~Tyurin, \emph{Reflective scattering at the lhc and
  two-scale structure of a proton},
  \href{https://doi.org/10.1209/0295-5075/129/31002}{\emph{Europhysics Letters}
  {\bfseries 129} (2020) 31002}.

\bibitem{Boreskov:2005ee}
K.G.~Boreskov, A.B.~Kaidalov, V.A.~Khoze, A.D.~Martin and M.G.~Ryskin,
  \emph{{The Partonic interpretation of reggeon theory models}},
  \href{https://doi.org/10.1140/epjc/s2005-02376-8}{\emph{Eur. Phys. J. C}
  {\bfseries 44} (2005) 523}
  [\href{https://arxiv.org/abs/hep-ph/0506211}{{\ttfamily hep-ph/0506211}}].

\bibitem{Glauber}
R.J.~Glauber, \emph{in lecture in theoretical physics}, {\emph{Vol. 1, edited
  by W. E. Brittin, L. G. Duham (Interscience, New York, 1959)} }.

\bibitem{Lipari2009}
P.~Lipari and M.~Lusignoli, \emph{Multiple parton interactions in hadron
  collisions and diffraction},
  \href{https://doi.org/10.1103/PhysRevD.80.074014}{\emph{Phys. Rev. D}
  {\bfseries 80} (2009) 074014}.

\bibitem{SIBYLL_2020}
F.~Riehn, R.~Engel, A.~Fedynitch, T.K.~Gaisser and T.~Stanev, \emph{Hadronic
  interaction model sibyll 2.3d and extensive air showers},
  \href{https://doi.org/10.1103/PhysRevD.102.063002}{\emph{Phys. Rev. D}
  {\bfseries 102} (2020) 063002}.

\bibitem{QGSJET}
{Ostapchenko, Sergey}, \emph{Qgsjet-iii model: physics and preliminary
  results}, \href{https://doi.org/10.1051/epjconf/201920811001}{\emph{EPJ Web
  Conf.} {\bfseries 208} (2019) 11001}.

\bibitem{GW_1960}
M.L.~Good and W.D.~Walker, \emph{Diffraction dissociation of beam particles},
  \href{https://doi.org/10.1103/PhysRev.120.1857}{\emph{Phys. Rev.} {\bfseries
  120} (1960) 1857}.

\bibitem{MP_1978}
H.I.~Miettinen and J.~Pumplin, \emph{Diffraction scattering and the parton
  structure of hadrons},
  \href{https://doi.org/10.1103/PhysRevD.18.1696}{\emph{Phys. Rev. D}
  {\bfseries 18} (1978) 1696}.

\bibitem{GLM_1999}
E.~Gotsman, E.~Levin and U.~Maor, \emph{Survival probability of large rapidity
  gaps in a three channel model},
  \href{https://doi.org/10.1103/PhysRevD.60.094011}{\emph{Phys. Rev. D}
  {\bfseries 60} (1999) 094011}.

\bibitem{Avsar_2007}
E.~Avsar, G.~Gustafson and L.~Lönnblad, \emph{Diffractive excitation in dis
  and pp collisions},
  \href{https://doi.org/10.1088/1126-6708/2007/12/012}{\emph{Journal of High
  Energy Physics} {\bfseries 2007} (2007) 012}.

\bibitem{Flensburg_2010}
C.~Flensburg and G.~Gustafson, \emph{Fluctuations, saturation, and diffractive
  excitation in high energy collisions},
  \href{https://doi.org/10.1007/jhep10(2010)014}{\emph{Journal of High Energy
  Physics} {\bfseries 2010} (2010) }.

\bibitem{Flensburg:2008ag}
C.~Flensburg, G.~Gustafson and L.~Lonnblad, \emph{{Elastic and quasi-elastic $p
  p$ and $\gamma^{*} p$ scattering in the Dipole Model}},
  \href{https://doi.org/10.1140/epjc/s10052-009-0868-7}{\emph{Eur. Phys. J. C}
  {\bfseries 60} (2009) 233} [\href{https://arxiv.org/abs/0807.0325}{{\ttfamily
  0807.0325}}].

\bibitem{Flensburg_2011}
C.~Flensburg, G.~Gustafson and L.~Lönnblad, \emph{Inclusive and exclusive
  observables from dipoles in high energy collisions},
  \href{https://doi.org/10.1007/jhep08(2011)103}{\emph{Journal of High Energy
  Physics} {\bfseries 2011} (2011) }.

\bibitem{Flensburg_2012}
C.~Flensburg, G.~Gustafson and L.~Lönnblad, \emph{Exclusive final states in
  diffractive excitation},
  \href{https://doi.org/10.1007/jhep12(2012)115}{\emph{Journal of High Energy
  Physics} {\bfseries 2012} (2012) }.

\bibitem{Gustafson_2015}
G.~Gustafson, L.~L\"onnblad, A.~Ster and T.~Cs\"org\H{o}, \emph{{Total,
  inelastic and (quasi-)elastic cross sections of high energy pA and gamma-A
  reactions with the dipole formalism}},
  \href{https://doi.org/10.1007/JHEP10(2015)022}{\emph{JHEP} {\bfseries 10}
  (2015) 022} [\href{https://arxiv.org/abs/1506.09095}{{\ttfamily
  1506.09095}}].

\bibitem{Bierlich:2016smv}
C.~Bierlich, G.~Gustafson and L.~L\"onnblad, \emph{{Diffractive and
  non-diffractive wounded nucleons and final states in pA collisions}},
  \href{https://doi.org/10.1007/JHEP10(2016)139}{\emph{JHEP} {\bfseries 10}
  (2016) 139} [\href{https://arxiv.org/abs/1607.04434}{{\ttfamily
  1607.04434}}].

\bibitem{Cudell_2009}
J.-R.~Cudell, E.~Predazzi and O.V.~Selyugin, \emph{New analytic unitarization
  schemes}, \href{https://doi.org/10.1103/PhysRevD.79.034033}{\emph{Phys. Rev.
  D} {\bfseries 79} (2009) 034033}.

\bibitem{Bhattacharya_2021}
A.~Bhattacharya, J.-R.~Cudell, R.~Oueslati and A.~Vanthieghem, \emph{Proton
  inelastic cross section at ultrahigh energies},
  \href{https://doi.org/10.1103/physrevd.103.l051502}{\emph{Physical Review D}
  {\bfseries 103} (2021) }.

\bibitem{Cudell:2003dz}
J.R.~Cudell and O.V.~Selyugin, \emph{{Saturation regimes at LHC energies}},
  {\emph{Czech. J. Phys.} {\bfseries 54} (2004) A441}
  [\href{https://arxiv.org/abs/hep-ph/0309194}{{\ttfamily hep-ph/0309194}}].

\bibitem{Hatlo_2005}
M.~Hatlo, F.~James, P.~Mato, L.~Moneta, M.~Winkler and A.~Zsenei,
  \emph{{Developments of mathematical software libraries for the LHC
  experiments}}, \href{https://doi.org/10.1109/TNS.2005.860152}{\emph{IEEE
  Trans. Nucl. Sci.} {\bfseries 52} (2005) 2818}.

\bibitem{Ostapchenko:2006vr}
S.~Ostapchenko, \emph{{On the re-summation of enhanced Pomeron diagrams}},
  \href{https://doi.org/10.1016/j.physletb.2006.03.026}{\emph{Phys. Lett. B}
  {\bfseries 636} (2006) 40}
  [\href{https://arxiv.org/abs/hep-ph/0602139}{{\ttfamily hep-ph/0602139}}].

\bibitem{Kolevatov:2012vu}
R.S.~Kolevatov and K.G.~Boreskov, \emph{{All-loop calculations of total,
  elastic and single diffractive cross sections in RFT via the stochastic
  approach}}, \href{https://doi.org/10.1063/1.4802135}{\emph{AIP Conf. Proc.}
  {\bfseries 1523} (2013) 137}
  [\href{https://arxiv.org/abs/1212.0691}{{\ttfamily 1212.0691}}].

\bibitem{PhysRevLett.127.062003}
{\scshape D0 collaboration\ifmmode\dagger\else\textdagger\fi{} and TOTEM
  Collaboration\ifmmode\ddagger\else\textdaggerdbl\fi{}} collaboration,
  \emph{Odderon exchange from elastic scattering differences between $pp$ and
  $p\overline{p}$ data at 1.96 tev and from $pp$ forward scattering
  measurements},
  \href{https://doi.org/10.1103/PhysRevLett.127.062003}{\emph{Phys. Rev. Lett.}
  {\bfseries 127} (2021) 062003}.

\bibitem{KMR_PRD_2018}
V.A.~Khoze, A.D.~Martin and M.G.~Ryskin, \emph{Elastic proton-proton scattering
  at 13 tev}, \href{https://doi.org/10.1103/PhysRevD.97.034019}{\emph{Phys.
  Rev. D} {\bfseries 97} (2018) 034019}.

\end{thebibliography}\endgroup
\end{document}